\documentclass[runningheads]{llncs}

\usepackage[T1]{fontenc}
\usepackage{url}  
\usepackage{tikz}
\usetikzlibrary{arrows.meta, positioning, shapes, calc}

\usepackage{graphicx}
\begin{document}

\title{STRIKE: A Structured Taxonomy of Cybercrime for Risk, Impact, Knowledge, and Evolution}
%
%

\author{  Melissa Pappy, Linh Nguyen, Suman Kumar,  Byungkwan Jung, \\ \and  Bernard Chen
}

\authorrunning{M.Pappy et al.}

\institute{Troy University, Troy AL 36082, USA \\
\email{\{mvenable150308, lnguyen93238, skumar, bjung, chenb \}@troy.edu}}

\maketitle              

\begin{abstract}
	Cybercrime has grown exponentially in both scale and sophistication, posing significant threats. As attack methods evolve rapidly, traditional classification schemes often fail to capture the complexity and diversity of modern threats. To address this gap, we introduce STRIKE, a Structured Taxonomy for Risk, Impact, Knowledge, and Emerging Threats, which provides a unified, multi-dimensional framework for categorizing cybercrimes. STRIKE spans both conventional and emerging domains, including ransomware, phishing, network intrusion, child sexual abuse material (CSAM), cryptojacking, deepfakes, and supply chain attacks. It organizes threats using criteria such as attack vectors, adversarial tactics, societal impact, detection techniques, and mitigation strategies. Alongside the taxonomy, we review recent advances in detection methodologies and present a response workflow to assist practitioners under active threat conditions. This work offers researchers, security professionals, and policymakers a practical foundation for threat analysis, comparative evaluation, and adaptive cyber defense.
    
\end{abstract}

\section{Introduction}
\label{intro}
Cyberspace is the new horizon, a globally interconnected domain controlled by machines for the exchange of information and communication among human beings. Crimes committed in cyberspace are collectively known as \emph{cybercrimes}, broadly defined as criminal acts perpetrated using computers, networks, or the Internet itself~\cite{r_2008}. A \emph{cybercriminal} is an individual who engages in or facilitates such activities. There are many kinds of cybercriminals~\cite{sabillon2016cybercrime}, including crackers, hackers, pranksters, career criminals, cyber terrorists, cyberbullies, and salami attackers, among others.

Given the widespread adoption of computer technology across all sectors of society from business operations to critical infrastructure the problem of information protection has become more urgent than ever. Early attacks relied on distributing malicious code via floppy disks and removable media, infecting systems one at a time. However, the Internet quickly evolved into the primary vector for mass infection and exploitation. Whether the mechanism is a computer virus, trojan, worm, or ransomware, attackers can now compromise thousands of systems within hours. High-profile incidents such as the \emph{WannaCry} ransomware outbreak (detected more than 213,000 times in 112 countries)~\cite{wannacry} have underscored the scale and speed at which cybercrime can disrupt economic activity, endanger democratic processes, and damage social trust. Reports consistently show that cybercrime has been growing exponentially, with Verizon’s Data Breach Investigations Reports documenting a continuous increase in both frequency and impact~\cite{verizon2017}.

In recent years, the threat landscape has further diversified. According to Europol’s 2023 Internet Organised Crime Threat Assessment~\cite{europol2023}, \emph{ransomware remains the most significant cyber threat}, fueled by the rise of \emph{Ransomware-as-a-Service (RaaS)} marketplaces and double extortion strategies, where attackers not only encrypt data but also threaten to leak sensitive information publicly. Simultaneously, phishing campaigns have become more sophisticated, often leveraging \emph{synthetic media} such as deepfake audio and video to deceive both individuals and organizations~\cite{brundage2018malicious}. The COVID-19 pandemic has dramatically expanded the attack surface by accelerating remote work, increasing cloud dependence, and normalizing digital transactions~\cite{kshetri2021covid}. Moreover, \emph{supply chain compromises}, such as the SolarWinds attack, have demonstrated that even well-defended networks remain vulnerable through trusted third-party vendors~\cite{solarwinds2021}.

Cybercrime today targets virtually every facet of society: individuals, small businesses, multinational corporations, and national governments. Criminals employ a wide array of tools, including AI-powered social engineering, cryptojacking malware, and advanced persistent threats (APTs)~\cite{nisioti2021intrusion,conti2022ransomware}. The World Economic Forum consistently ranks cybercrime among the most pressing global risks, citing its potential to undermine economic stability and national security~\cite{wef2023}.

In response to the expanding landscape of cyber threats targeting individuals, businesses, and critical infrastructure, building on prior work~\cite{chandra2020taxonomy,donalds2019ontology},  this paper introduces STRIKE: a Structured Taxonomy for Risk, Impact, Knowledge, and Emerging threats. The STRIKE framework offers a comprehensive classification of modern cybercrimes by integrating traditional domains (e.g., phishing, intrusion, ransomware) with rapidly evolving threats such as deepfake-enabled fraud, cryptojacking, and supply chain attacks. Each threat category is examined through a multi-dimensional lens that is its delivery method, impact profile, attacker motivation, and defensive countermeasures. By aligning taxonomy development with current detection, mitigation, and policy needs, STRIKE serves as both an analytical model and a practical guide for cybersecurity practitioners, researchers, and policy-makers. Its extensible structure also enables continuous adaptation to new attack vectors and evolving threat actor behaviors. By consolidating insights from technical, legal, and societal perspectives, this taxonomy aims to support researchers, practitioners, and policymakers in developing more resilient and informed cybersecurity strategies.

The remainder of the paper is organized as follows: Section~\ref{Existing} reviews key recent contributions in the field of cybercrime research and outlines how our approach differs from existing taxonomies and frameworks. Section~\ref{Methods} describes the methodology followed to develop STRIKE. Section~\ref{types} introduces the proposed STRIKE taxonomy, detailing various cybercrime categories along with their impacts and associated detection techniques. Furthermore, we examine commonly adopted mitigation strategies and discuss broader issues related to internet governance and cybersecurity policy in the same section. Finally, Section~\ref{future} concludes the paper and outlines potential future research directions aimed at strengthening cyber resilience in the face of evolving digital threats.

\section{Existing work}
\label{Existing}
Over the past two decades, researchers have proposed a variety of taxonomies, ontologies, and detection frameworks to classify cybercrime and support defensive strategies. This section summarizes the most notable contributions and highlights how the proposed work differs from existing literature.

Development of a theory-based taxonomy for cybercrime and the need for clear definitions and standards for measurement and management are presented in~\cite{chandra2020taxonomy}. Chandra and Snowe describe the building blocks of their taxonomy, including mutual exclusivity, structure, exhaustiveness, and well-defined categories, which form a robust theoretical foundation. In addition, their work discusses potential implications in the realms of management, reporting, disclosure, governance, regulation, and the judiciary. However, while conceptually rigorous, this approach primarily emphasizes definitional clarity rather than the integration of contemporary detection methodologies or practical response workflows.

The need for a universally agreed-upon classification scheme to enhance understanding of cybercrime and to facilitate legal and policy responses is tackled in~\cite{DONALDS2019403}. Donalds and Osei-Bryson introduce a cybercrime ontology that incorporates multiple perspectives and aims to offer a more holistic viewpoint for classification. Their work advances the field by acknowledging the diversity of stakeholder perspectives and the need for a unified conceptual framework. Nonetheless, it predates the recent shifts in the threat landscape, such as ransomware-as-a-service and deepfake-enabled phishing, and does not explicitly link taxonomy categories to detection and mitigation strategies.

Al-Khater et al.~\cite{9146148} conducted a comprehensive review of cybercrime detection and prevention techniques, exploring various types of cybercrimes, their impacts (e.g., privacy breaches, security violations, business losses, financial fraud), and the strategies commonly employed by attackers. While this review provides valuable insights into detection methods and recommendations for more effective models, it does not propose a unified taxonomy that links threat categories to countermeasures or response workflows. 

More recent surveys illustrate the growing complexity of the field. Conti et al.~\cite{conti2022ransomware} examined the evolution of ransomware, highlighting the rise of double extortion and the increasing importance of behavior-based detection approaches. Nisioti et al.~\cite{nisioti2021intrusion} reviewed machine learning techniques for intrusion detection, emphasizing both their potential and their limitations, including challenges with false positives and dataset biases. Kshetri~\cite{kshetri2021covid} analyzed the surge in cybercrime driven by the COVID-19 pandemic and the rapid expansion of remote work. Brundage et al.~\cite{brundage2018malicious} discussed the malicious use of artificial intelligence, including deepfake-enabled social engineering and automated reconnaissance. Recently, there has been a rapid evolution of ransomware evasion strategies~\cite{raman2024ransomware}, deep learning-based intrusion detection~\cite{smith2024intrusion}, and the critical impact of deepfake-enabled social engineering~\cite{lee2024deepfake}. Ali and Zhang~\cite{ali2025phishing} have documented the surge in AI-powered phishing attacks, while Chen and Gupta~\cite{chen2024supplychain} have classified emerging supply chain threats in IoT environments. Rodriguez and Ahmed~\cite{rodriguez2025cybercrime} provide a global overview of new cybercrime categories, motivations, and countermeasures. 

Several international reports also contribute important perspectives. The Europol Internet Organised Crime Threat Assessment (IOCTA)~\cite{europol2023} and the World Economic Forum’s Global Risks Report~\cite{wef2023} document the escalation of ransomware-as-a-service marketplaces, cryptojacking, supply chain compromises, and synthetic media attacks.

Despite the progress made by these contributions, existing work exhibits several limitations. Taxonomies and ontologies are often static and may not adequately capture emerging threats such as deepfake fraud, cryptojacking, and IoT supply chain compromises.
 Many frameworks stop at conceptual categorization and do not integrate detection methodologies, response workflows, or mitigation strategies in a practical manner. There is limited focus on accessibility for multidisciplinary stakeholders, including policymakers, educators, and non-technical people. While prior works have offered taxonomies focused on cyberattack vectors or technical vulnerabilities, they often emphasize technical classification or system-layer attacks alone. STRIKE differs from existing efforts by explicitly incorporating impact analysis, adversary behavior, and mitigation strategies as integral classification dimensions. For example, unlike Chandra and Snowe’s theory-driven model~\cite{CHANDRA2020100467} or Donalds and Osei-Bryson’s ontology~\cite{DONALDS2019403}, STRIKE integrates new and underrepresented cybercrimes such as CSAM, synthetic identity fraud, and deepfake-based impersonation—within a unified framework. Moreover, STRIKE bridges academic insight with operational relevance by structuring attack profiles alongside their associated risks, social impact, and current defense practices, filling an important gap between theoretical taxonomy and applied cybersecurity defense.

\section{Methodology}
\label{Methods}
This study adopts a multi-phase methodology to construct and validate the proposed STRIKE (Structured Taxonomy for Risk, Impact, Knowledge, and Emerging Threats) framework for cybercrime classification. The methodology comprises four key stages:

\subsection{Literature Survey and Source Selection}
A structured literature review was conducted to capture the current landscape of cybercrime detection, mitigation, and classification. Peer-reviewed journals, conference proceedings, industry whitepapers, technical reports, and governmental publications were surveyed using databases such as IEEE Xplore, ACM Digital Library, SpringerLink, and arXiv. Search terms included ransomware detection, phishing attacks, cybercrime taxonomy, deepfake detection, and child safety online. Sources were selected based on relevance, recency (primarily 2017--2024), citation strength, and practical significance.

The survey methodology followed a targeted thematic approach. Literature was categorized by cybercrime domain, e.g., ransomware, phishing, CSAM, intrusion, cryptojacking, and analyzed across consistent dimensions: attack vectors, adversarial techniques, targets, impact, detection methods, and mitigation strategies. Preference was given to comprehensive surveys, technical frameworks, and meta-analytical studies that addressed emerging threats and novel defense strategies. Representative works include~\cite{conti2022ransomware}, Donalds \& Osei-Bryson~\cite{DONALDS2019403}, and Chandra \& Snowe~\cite{CHANDRA2020100467}. This process enabled a comparative synthesis of threats and informed the construction of the STRIKE taxonomy.

\subsection{STRIKE Construction}
Cybercrime characteristics were systematically extracted and organized across several dimensions, including attack techniques and delivery vectors, target entities and adversary profiles, impact domains spanning technical, financial, psychological, and societal consequences, detection and mitigation strategies, and relevant legal, policy, and governance contexts. These dimensions were synthesized into two comprehensive taxonomy tables: one that consolidates detection, mitigation, and prevention strategies across cybercrime categories, and another that provides detailed profiles of each attack type, highlighting methods, vectors, targets, and threat actors.

\subsection{Mapping and Validation}
To ensure completeness and operational alignment, the STRIKE taxonomy was mapped against established cybersecurity frameworks such as the NIST Cybersecurity Framework, MITRE ATT\&CK, and Europol IOCTA reports. Threat categories were also cross-validated with industry threat intelligence, including the Verizon DBIR and GAO's SolarWinds incident report.

\subsection{Analysis and Profiling}
Real-world attack scenarios were integrated to contextualize each cybercrime category. For each category, typical adversary behaviors, technical indicators, and known response workflows were examined. Where applicable, illustrative examples and empirical findings from prior literature were used to demonstrate the taxonomy's practical relevance and extensibility.

\section{STRIKE: A Structured Taxonomy of Cybercrime }
\label{types}

\begin{figure}[htp]
    \centering
    \includegraphics[width=1.0\textwidth, height=2.5\textheight, keepaspectratio]{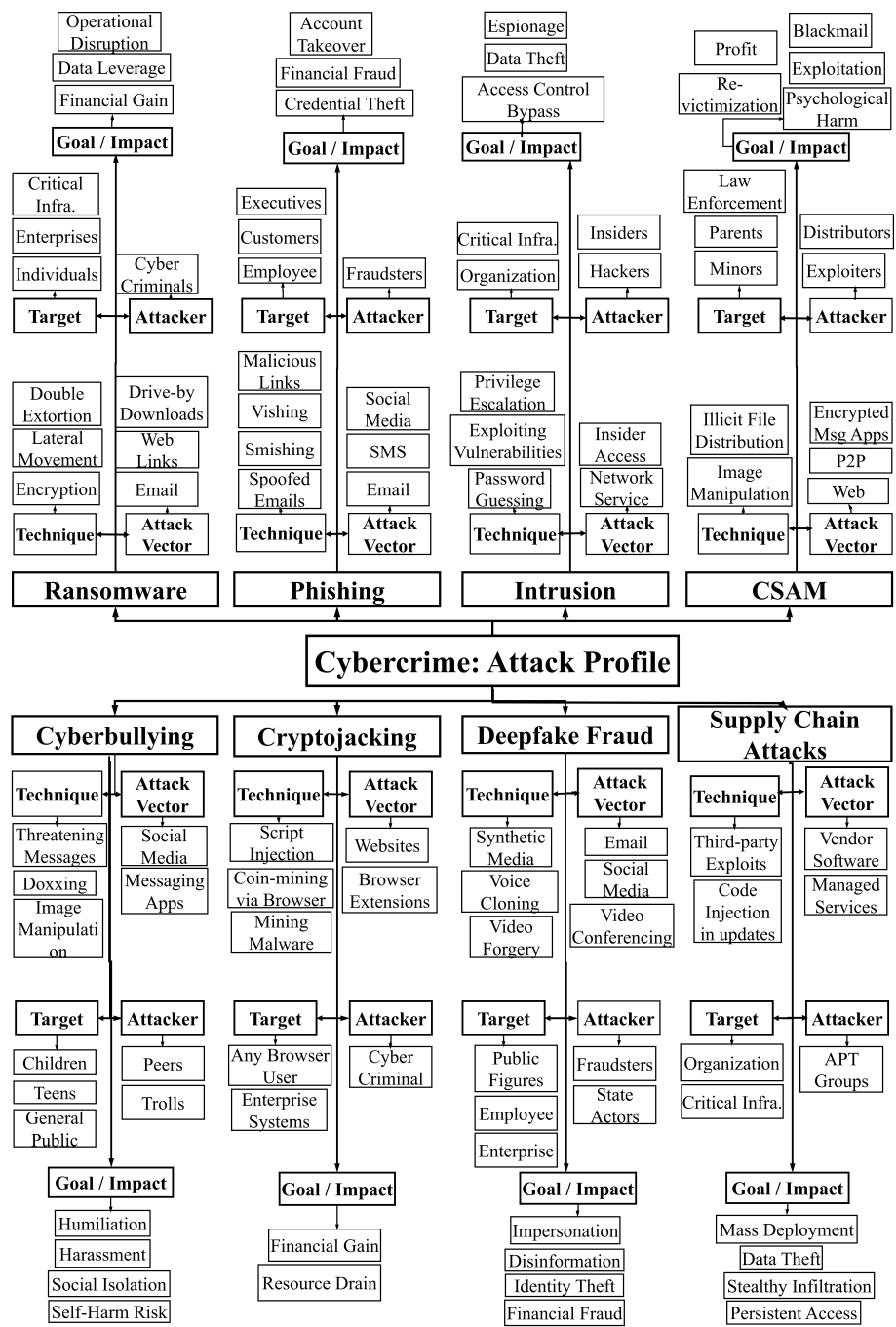} 
    \caption{Cybercrime: Attack Profile}
    \label{tab:cybercrime-profile}
\end{figure}

Figure~\ref{tab:cybercrime-profile} and Table~\ref{tab:cybercrime_full} provide a unified view of the cybercrime landscape. The figure outlines attack profiles by connecting techniques, vectors, attackers, targets, and impacts, while the table complements this by detailing detection, mitigation, and prevention strategies for each category. Together, they highlight both shared patterns and unique features of threats ranging from traditional attacks like ransomware and phishing to emerging risks such as cryptojacking and deepfake-enabled fraud, offering a practical foundation for threat modeling and defense planning.

\subsection{Ransomware}
Ransomware is malicious software that encrypts or locks a victim's files and systems, demanding payment—typically in cryptocurrency in exchange for the password that can decrypt or unlock the system~\cite{conti2022ransomware}. Historically, ransomware emerged as scareware or locker ransomware targeting individuals. Over the past decade, it has evolved into a sophisticated ecosystem with extensive underground infrastructure, giving rise to \emph{Ransomware-as-a-Service} (RaaS) operations~\cite{raman2024ransomware}. In these models, ransomware developers lease their tools to others who conduct attacks and then split ransom proceeds among themselves. This Software-as-a-Service model has made it significantly easier for cybercriminals to launch attacks, fueling exponential growth in attacks.

\begin{table*}[htp]
\centering
\caption{Cybercrime: Detection, Mitigation, and Prevention}
\label{tab:cybercrime_full}
\scriptsize
\begin{tabular}{|p{1.7cm}|p{3.2cm}|p{3cm}|p{3.2cm}|p{0.6cm}|}
\hline
\textbf{Category} & \textbf{Detection} & \textbf{Mitigation} & \textbf{Prevention} & \textbf{Key Ref.} \\
\hline
Ransomware & File system monitoring, memory forensics, entropy and I/O pattern analysis, honeypots, anomaly detection & System isolation, decryption tools, restoring from clean backups, threat response teams & Offline and cloud backups, software patching, access control, user training & \cite{conti2022ransomware}, \cite{raman2024ransomware}, \cite{kharraz_robertson_balzarotti_bilge_kirda_2015}, \cite{o'gorman_mcdonald_2012}, \cite{smith2024intrusion}, \cite{chen2024supplychain} \\
\hline
Phishing & Email header analysis, link inspection, NLP classifiers, blacklist and reputation scoring, browser plugins & Blacklisting, warning systems, credential resets, alerting security teams & Anti-phishing training, spam filters, multi-factor authentication (MFA), email protocol enforcement (DMARC, SPF, DKIM) & \cite{ali2025phishing}, \cite{basnet2021deep}, \cite{hong2012state}, \cite{parsons2015phishing}, \cite{lee2024deepfake} \\
\hline
Network Intrusions & Signature/anomaly-based IDS, flow analysis, log correlation, kill chain modeling & Quarantine endpoints, patch vulnerabilities, credential revocation, incident response & Segmentation, least privilege, frequent audits, zero-trust architecture & \cite{scarfone2007guide}, \cite{liao2013intrusion}, \cite{xu2021survey}, \cite{hutchins2011kill}, \cite{bace2000intrusion} \\
\hline
Child Sexual Abuse Material (CSAM) & Perceptual hashing (PhotoDNA), CNN-based image recognition, facial landmark estimation, keyword filtering & Image takedown, URL/IP blacklisting, law enforcement reporting, suspect tracing & Platform moderation, legal frameworks, child protection education, collaboration with NGOs & \cite{microsoft2019photodna}, \cite{wang2022csam}, \cite{TwotierChild2016}, \cite{interpol2021}, \cite{lee2024deepfake} \\
\hline
Cyberbullying & Sentiment analysis, NLP-based toxicity detection, deepfake detection, complaint systems & Post removal, account suspension, psychological support, auto-moderation & Awareness campaigns, digital literacy education, active monitoring, community standards & \cite{salawu2020survey}, \cite{xiang2021deepfake}, \cite{smith2014cyberbullying} \\
\hline
Cryptojacking & Resource usage anomalies, web mining signature analysis, endpoint alerts, script auditing & Process termination, cleanup tools, block mining pools/IPs, extension removal & Browser hardening, JS filtering, no-script policies, endpoint protection platforms & \cite{eskandari2018survey}, \cite{tahir2021cryptojacking}, \cite{hong2021survey} \\
\hline
Deepfake-Enabled Fraud & Video/audio forensics, liveness detection, ML anomaly detection in media, watermark tracing & Content takedown, account review, source verification, forensic review & Blockchain-based media validation, public awareness, regulation, detection tools for journalists & \cite{mirsky2021creation}, \cite{tolosana2020deepfakes}, \cite{chesney2019deepfakes}, \cite{verdoliva2020media} \\
\hline
Supply Chain Attacks & Behavior profiling, SBOM diffing, code signing, vendor intelligence & Disconnect vulnerable links, revoke certificates, block updates, notify partners & Secure development lifecycle (SDLC), vendor audits, inventory monitoring, endpoint protection & \cite{boyens2020nist}, \cite{chen2024supplychain}, \cite{solarwinds2021}, \cite{tan2018ccleaner} \\
\hline
\end{tabular}
\end{table*}

Ransomware can be delivered through email attachments or malicious links that trick the user into downloading the ransomware software~\cite{o'gorman_mcdonald_2012}. Often, ransomware is present on compromised websites, also known as drive-by downloads, which infect users when they visit these sites containing hidden iframes or exploit unpatched vulnerabilities in the system. Once installed, the malware executes and takes the system hostage~\cite{kharraz_robertson_balzarotti_bilge_kirda_2015}.

Ransomware is of two types: \emph{Locker Ransomware} and \emph{Crypto Ransomware}\cite{pathak_nanded_2016}. Locker ransomware locks the desktop or the user interface, preventing access but leaving files intact. This form is generally less destructive, since systems can be restored after removing the malware. In contrast, crypto ransomware silently searches for valuable files and encrypts them without the user's knowledge, rendering them inaccessible and forcing the user to pay a ransom\cite{pathak_nanded_2016}.

In modern ransomware attacks, attackers exfiltrate data before encryption and threaten public release if payment is not made~\cite{conti2022ransomware}. This tactic increases leverage over victims and has contributed to ransomware becoming one of the most financially damaging forms of cybercrime~\cite{europol2023}. Often, the victim does not report the attack to the public or even to authorities, fearing it might damage their reputation and cause more harm than simply paying the ransom.

\subsubsection{Impact}
Ransomware attacks play on the fear of victims, who are often so anxious about losing critical data or facing reputational harm that they pay without verifying whether recovery is possible. Payments are usually demanded in untraceable cryptocurrencies such as Bitcoin~\cite{o'gorman_mcdonald_2012}. In many cases, even after payment, the files remain unrecoverable, resulting in permanent data loss.
Studies show that approximately 2.9\% of victims ultimately pay the ransom~\cite{o'gorman_mcdonald_2012}. Attackers typically demand amounts ranging from \$50 to \$200, and with sufficient scale, can earn hundreds of thousands of dollars monthly. For example, one estimate suggests that an attacker could make up to \$394{,}400 in a single month, factoring in money laundering and conversion fees~\cite{castro2017}. 

Beyond financial losses, ransomware has profound psychological impacts, including anxiety, depression, and social alienation~\cite{kshetri2021covid,rodriguez2025cybercrime,wef2023}. Incidents affecting hospitals, schools, and critical infrastructure have demonstrated that ransomware can also pose serious risks to public health and safety.

\subsubsection{Detection, Mitigation and Prevention}

Detecting ransomware remains a challenging task due to continuous innovation in evasion techniques. Various detection strategies have been proposed, including file system and I/O monitoring, which identify ransomware by extracting distinctive access patterns characterized by tuples of process, file, operation type, and entropy~\cite{kharraz_robertson_balzarotti_bilge_kirda_2015,kharraz_robertson_arshad_kirda_mulliner_2016}. Locker ransomware can also be detected through desktop monitoring techniques that capture and analyze screenshots for ransom messages. Graph-based approaches construct file relationship graphs using metrics like Jaccard similarity and locality-sensitive hashing to classify malicious files~\cite{Tamersoy:2014:GAL:2623330.2623342}. Memory and process analysis assist in detecting in-memory encryption activities~\cite{raman2024ransomware}, while machine learning methods analyze system call sequences, entropy, and behavioral features to detect anomalies~\cite{smith2024intrusion}. Honeypots and deception-based systems deploy decoy files or environments to observe and detect early ransomware activity~\cite{conti2022ransomware}. Mitigation and prevention demand a layered strategy including endpoint protection, user awareness training, behavioral monitoring, frequent backups, and incident response planning. Additional measures, such as timely patching, restricting administrative privileges, and enforcing multi-factor authentication, help minimize exposure to ransomware attacks.

\subsubsection{Emerging Challenges}
Ransomware continues to adapt through obfuscation, polymorphism, and in-memory execution, which hinder detection and forensics~\cite{conti2022ransomware,raman2024ransomware}. It exploits trusted tools like PowerShell and targets supply chains, as seen in the Kaseya breach~\cite{chen2024supplychain}. Double extortion tactics and Ransomware-as-a-Service have accelerated growth and impact~\cite{europol2023,raman2024ransomware} since one does not require deep technical skills. Attacks on critical infrastructure underscore the need for layered defenses, proactive monitoring, and robust response plans~\cite{kshetri2021covid}. These approaches rely on supervised learning and therefore, lacks the ability to detect new types of Ransomwares.

\subsection{Computer and Network Intrusions}
Intrusion is an act by a user of an information system who is not legally authorized to take a particular action. This user is referred to as an intruder. Intrusions can be performed by external attackers exploiting vulnerabilities in network services or by insiders exceeding their legitimate privileges~\cite{sharma2020intrusion}. Such attacks range from simple password guessing to sophisticated advanced persistent threats (APTs) that remain undetected for extended periods~\cite{sharma2023advanced}.

\subsubsection{Impact}

Computer files, databases, networks, and Internet-based applications have gradually become critical assets for organizations. When these assets are attacked or compromised, data integrity and privacy are threatened, business operations are disrupted, and customer trust may be lost—sometimes resulting in severe financial and reputational damage~\cite{ponemon2023cost}. In critical infrastructure contexts, intrusions can interrupt essential services such as energy grids, healthcare systems, and transportation networks~\cite{gandotra2014survey}.

\subsubsection{Detection, Mitigation and Prevention}
Intrusion Detection (ID) is a type of security management system for computers and networks. An IDS gathers and analyzes information from multiple sources to detect security breaches. Typically, IDSs are classified by their monitoring location and detection strategy:

\begin{itemize}
    \item Host-Based IDS (HIDS) monitors the activities of specific machines, collecting data such as audit logs, configuration files, and system calls~\cite{bace2000intrusion}.
    \item Network-Based IDS (NIDS) captures and analyzes packets traversing the network, often using deep packet inspection or flow analysis~\cite{NetSTAT_A_Network_based_Intrusion_Detection_Approach_Vigna_Kemmmerer}.
    \item Hybrid IDS combines host and network data and may take proactive measures (e.g., blocking traffic) or simply alert administrators.
    \item Advanced Persistent Threats (APTs) involve prolonged, targeted campaigns against specific organizations. Detection often relies on correlating events over time, combining multiple data sources, and using kill-chain analysis frameworks~\cite{hutchins2011kill,xu2021survey}. Graph-based anomaly detection and threat intelligence integration are increasingly important in uncovering APT activity.
\end{itemize}

There are two main IDS techniques. Signature-based IDS searches for known patterns or byte sequences indicative of attacks~\cite{scarfone2007guide}. Anomaly detection establishes a baseline of normal system or network behavior and flags deviations~\cite{Computer_Network_Intrusion_Detection_zhang_1996}. Modern systems increasingly combine both approaches in hybrid and ensemble methods~\cite{liao2013intrusion}.

The effectiveness of mitigation approaches depends on multiple layers of protection. Network segmentation limits lateral movement, while multi-factor authentication reduces the risk of credential compromise. Other important measures include regular software patching, integrating threat intelligence, and enforcing the principle of least privilege to prevent unauthorized access.

\subsubsection{Emerging Challenges}
Proliferation of end-to-end encryption reduces packet visibility, creating a major challenge in intrusion detection~\cite{zhao2019iot}. Attackers nowadays deceptively use the combination of malicious actions with legitimate processes (living-off-the-land techniques) ~\cite{barr2021survivalism,ongun2021living}. Modern intrusion attacks include AI, making it adaptive to detection techniques ~\cite{shokri2021privacy}. The widespread use of untested and vulnerable IoT devices significantly expands the attack surface, especially when there are many autonomous systems combined to participate in the ecosystem, providing some services.

\subsection{Phishing and Social Engineering}
Phishing is a form of online identity theft in which attackers impersonate legitimate entities to trick victims into revealing sensitive information, installing malware, or transferring funds~\cite{hong2012state}. Also known as ``carding'' or ``brand spoofing,'' phishing attacks frequently target individuals and organizations through email, SMS (smishing), and voice calls (vishing)~\cite{www.justice.gov_2006}. Modern campaigns increasingly leverage AI-generated content and deepfake audio or video to enhance credibility~\cite{lee2024deepfake}. There are four main types of phishing attacks: spear phishing, which targets specific individuals or organizations; whaling, which aims at senior executives or high-profile figures; clone phishing, which duplicates legitimate emails but replaces links or attachments with malicious content; and vishing and smishing, which rely on voice and SMS-based campaigns.

A typical phishing attack proceeds in four stages~\cite{chawla_chouhan_2014}. First, the attacker creates a fraudulent website that closely mimics a legitimate service, complete with realistic domain names, web pages, and branding. Second, large volumes of deceptive emails or messages are distributed to potential victims. Third, unsuspecting users are lured to the fake site and prompted to enter credentials, payment details, or other sensitive information. Finally, attackers use the stolen data to commit fraud or gain unauthorized access to accounts. Techniques such as URL obfuscation, compromised legitimate accounts, and malware-laden attachments are commonly employed to improve success rates~\cite{ali2025phishing}.

\subsubsection{Impact}
Studies have shown that over 15{,}000 phishing websites are created daily, with 95\% automatically generated by attacker infrastructure~\cite{sans_institute_infosec_reading_room_2007}. The impact of phishing extends beyond privacy breaches to include credential theft, unauthorized access, and identity fraud. These attacks often lead to fraudulent transactions~\cite{6585241}, financial losses, malware infections such as ransomware, and significant reputational damage. Phishing remains the most common initial access vector for ransomware and costs organizations billions of dollars annually~\cite{ali2025phishing}.

\subsubsection{Detection, Mitigation and Prevention}
Detection strategies fall into two main categories~\cite{chhikara_dahiya_garg_rani_2013}. Server-based detection includes brand monitoring to identify fraudulent websites, track abnormal activity, and aggregate logs to detect coordinated attacks. Detected phishing domains are added to centralized blacklists to block access. Client-based detection focuses on user devices, where browsers and email clients load URL blacklists, analyze message content with machine learning~\cite{basnet2021deep}, and monitor information flow for suspicious behavior~\cite{DAVIDSON20217}. Mitigation involves updating blacklists, disabling compromised accounts, and removing malicious content. Whereas prevention relies on user education, as awareness programs significantly reduce click-through rates and improve resilience against phishing~\cite{parsons2015phishing}.

\subsubsection{Emerging Challenges}

AI-generated emails and deepfake audio are increasingly convincing~\cite{lee2024deepfake}. There is an increase in Multi-channel campaigns targeting users across email, SMS, and social media~\cite{yang2019phishing,kaur2025survey}. There have been rise of malware-as-a-service offerings~\cite{patsakis2024malware}.

\subsection{Child Sexual Abuse Material (CSAM)}
Child Sexual Abuse Material (CSAM) refers to any visual representation depicting the sexual exploitation of minors, including photographs, videos, digital images, and other media~\cite{europol2023}. CSAM includes both real and simulated explicit conduct involving a minor, as well as any depiction of a minor’s sexual parts for primarily sexual purposes~\cite{introduction_refining_child_pornography_law_crime_language_and_social_consequences_2016,what_legally_makes_it_child_pornography?}. Even images that do not depict sexual activity, such as nudity presented in a sexualized context, may be classified as illegal. Visual material can include photographs, videos, digital downloads, undeveloped film, and electronic files.

\subsubsection{Impact}
The impact of CSAM on victims is severe and enduring. Long-term psychological trauma, social stigmatization, re-victimization each time the material is shared or accessed, and barriers to recovery due to the permanence of online distribution are just a few of the consequences. From a societal perspective, CSAM fuels a market for child exploitation, normalizes predatory behavior, and contributes to broader criminal networks~\cite{europol2023}. Research suggests that the consumption of such material can have a corrosive effect on offenders. Viewing CSAM may prime individuals to escalate from passive consumption to hands-on offending, normalize deviant sexual interests, and increase desensitization to harm~\cite{Child_Pornography_Research_Paper_2011}.

\subsubsection{Detection, Mitigation and Prevention}

Combating CSAM requires a combination of technical safeguards, legal frameworks, and international cooperation. Detection relies on hash databases maintained by organizations such as NCMEC and Interpol, with tools like PhotoDNA generating perceptual hashes to identify altered images~\cite{microsoft2019photodna}. Deep learning models, including convolutional neural networks, can classify content based on facial features, body proportions, and contextual clues~\cite{wang2022csam}. Some systems estimate age using facial landmarks~\cite{2015child,7574915}, while two-tier representations improve robustness against manipulation~\cite{TwotierChild2016}.

Mitigation involves platform monitoring and reporting workflows that automate scanning, flag content, and coordinate removal. Cross-border law enforcement collaboration helps identify offenders and rescue victims~\cite{interpol2021}. Preventative measures include proactive blocking of uploads, strong platform policies, and user reporting mechanisms. Rigorous laws and severe penalties—such as imprisonment, sex offender registration, and fines—serve as deterrents~\cite{introduction_refining_child_pornography_law_crime_language_and_social_consequences_2016}.

\subsubsection{Emerging Challenges}
The proliferation of online platforms, encrypted messaging applications, peer-to-peer networks, and anonymizing technologies has increased circulation and complicated enforcement~\cite{interpol2021}. Offenders increasingly rely on hidden services and dark web markets. Deepfake technology enables the creation of synthetic CSAM, complicating legal and technical responses~\cite{lee2024deepfake}, making it extremely challenging to tackle since it is now easy to create. It is worth noting that machine learning based detection accuracy varies widely (between 70\% and 85\%) and false positives remain a challenge~\cite{TwotierChild2016}.
\subsection{Cyberbullying}

Cyberbullying is the use of electronic communication technologies to intimidate, harass, or harm individuals, typically through repeated aggressive behavior~\cite{smith2014cyberbullying}. The U.S. Department of Health and Human Services defines cyberbullying as willful and repeated harm inflicted using electronic devices and communication tools such as social networks, text messaging, and websites~\cite{Hindu2014,gov1}. While the term generally applies to minors, similar behaviors targeting adults are referred to as cyberharassment or cyberstalking~\cite{smith2014cyberbullying,stopURL}. Cyberbullying is most prevalent among children and adolescents ages 6--18~\cite{Gina2014,gov3}, although adults increasingly experience online abuse on social media and messaging platforms. Cyberbullying includes sending threatening or humiliating messages, spreading rumors or manipulated images, impersonation to damage reputation, and orchestrating campaigns to isolate individuals. The proliferation of smartphones, instant messaging apps, and social networking platforms has amplified the reach and persistence of online harassment~\cite{barlett2019cyberbullying}.

\subsubsection{Impact}

Cyberbullying has severe psychological, emotional, and social consequences. Anxiety, depression, and decreased self-esteem can lead to disengagement from regular or essential activities. It often causes relationship challenges and increases the risk of self-harm and suicidal tendencies~\cite{kowalski2014bullying,Patchin3}. Victims are twice as likely to attempt suicide compared to non-victims~\cite{Patchin3}. Emotional consequences frequently include feelings of frustration, anger, sadness, helplessness, and vengefulness~\cite{Hindu2007}.

\subsubsection{Detection, Mitigation, and Prevention}

Detection of cyberbullying relies on a combination of manual and automated approaches. Natural language processing models automatically flag abusive or harassing text by analyzing sentiment, toxicity, and linguistic cues~\cite{salawu2020survey}. Computer vision techniques detect manipulated or explicit imagery used to embarrass or threaten victims, including deepfakes~\cite{xiang2021deepfake}. Platforms also monitor user behavior to identify patterns consistent with harassment, such as repeated targeting or coordinated attacks. Most social media services and messaging apps provide tools to report abuse, though effectiveness depends on enforcement and moderation workflows~\cite{hinduja2014cyberbullying}.

Mitigation includes support services such as counseling and helplines to address psychological harm. Clear platform policies and community standards should be consistently enforced~\cite{smith2014cyberbullying}. Prevention requires collaboration among parents, educators, and platform providers. Discouraging bullying, implementing enforcement policies, and integrating anti-bullying programs into schools can help reduce incidents. At the family level, promoting safe online behavior, monitoring online activities when appropriate, and maintaining open communication are essential. Coordinated efforts across communities, technology platforms, and policymakers, combined with early intervention and education, can significantly reduce the likelihood and impact of cyberbullying~\cite{Notar2013}.

\subsubsection{Emerging Challenges}
Surveys estimate that up to 35\% of adolescents have experienced online harassment~\cite{barlett2019cyberbullying}. The COVID-19 pandemic further increased incidents due to greater reliance on digital communication~\cite{Hindu2014}. Deepfake technologies are enabling more sophisticated forms of defamation and harassment~\cite{xiang2021deepfake}. Cross-border jurisdictional issues often limit the enforcement of protective laws. Cyberbullying is exacerbated by the ease of anonymity, widespread use of end-to-end encryption, and the difficulty of removing harmful content once it has been shared~\cite{gov1}.

\subsection{Supply Chain Attacks}
Supply chain attacks are intended to create disruption across the end-to-end delivery of goods and services. The supply chain includes suppliers, providers, third-party service vendors, and their supporting infrastructures. This type of attack exploits the trust ecosystem among multiple entities in the chain. It can target hardware or software and may involve anything from malicious software updates to backdoor insertion at any point in the supply chain infrastructure~\cite{boyens2020nist}.

\subsubsection{Impact of Supply Chain Attacks}

Malicious code insertion can disrupt operations and spread rapidly across the many organizations involved~\cite{solarwinds2021}. Trojan horses are often installed under the guise of legitimate software updates, affecting large numbers of users~\cite{tan2018ccleaner}. Ransomware attacks can impact systems ranging from accounting platforms to critical databases~\cite{greenberg2018notpetya}. Major consequences include compromise of government and enterprise networks, theft of intellectual property and sensitive data, operational disruption, reputational damage, and, in some cases, significant financial losses and social unrest.

\subsubsection{Detection and Mitigation}

Detection and mitigation techniques include code signing, authentication, and strict control of software changes~\cite{boyens2020nist}, as well as comprehensive software update tracking~\cite{boyens2020nist}. End-to-end monitoring and anomaly detection help identify suspicious activities across the supply chain~\cite{chen2024supplychain}. Because modern supply chains are highly complex and global in scope, securing them requires cross-industry collaboration, regulatory frameworks, and continuous oversight. Given the number of stakeholders involved, mitigation strategies increasingly rely on autonomous trust mechanisms among entities. Blockchain has emerged as a promising solution to facilitate secure, verifiable transactions and improve supply chain transparency~\cite{xu2021blockchain,agarwal2022blockchain}.

\subsection{Cryptojacking}

Cryptojacking is one of the most modern forms of cybercrime, in which attackers use victim\'s computing resources to mine cryptocurrency without their knowledge~\cite{eskandari2018survey}. A report by Symantec revealed that cryptojacking incidents grew by 8,500\% between 2017 and 2018 as attackers shifted from ransomware to stealthy mining~\cite{hong2021survey}. Attackers embed mining scripts in websites, exploit software vulnerabilities, or distribute malicious browser extensions~\cite{ketharanathan2019cryptojacking}, remaining hidden to maximize profit over time. Typically, this crime involves three entities: the third-party website, the user who visits it, and the attacker. In browser-based mining, attackers inject JavaScript code into web pages, causing visitors' browsers to mine cryptocurrency while they browse~\cite{eskandari2018survey}. Cryptojacking can also be achieved by installing malicious software directly on the victim's machine~\cite{conti2022ransomware}. Additionally, attackers may compromise cloud environments to deploy mining software and conduct large-scale mining operations~\cite{hong2021survey}.

\subsubsection{Impact}

Cryptojacking increases CPU and GPU usage, which reduces performance, disrupts normal device operation, and shortens hardware lifespan~\cite{meland2019experimental}. In cloud environments, it can lead to significantly higher electricity costs, increased resource consumption, and rising maintenance expenses. In addition, there are ramifications in a variety of critical sectors such as banking and finance ~\cite{kshetri2024cryptoran}. Recent study~\cite{adeniran2024dissecting} found that over half the website surveyed remained malicious, hosting mining scripts like CoinHive and Webminepool. The study revealed that cryptojacking infrastructure often overlaps with other threats, including phishing and malware distribution.

\subsubsection{Detection and Mitigation}

Detection strategies for cryptojacking include monitoring resource utilization, where unusual CPU or GPU activity, especially on idle systems, is a common indicator. Machine learning algorithms are increasingly useful for detecting such anomalies~\cite{tahir2021cryptojacking}. Monitoring network traffic helps identify outbound connections to mining pools and suspicious DNS requests~\cite{ketharanathan2019cryptojacking}. Signature-based detection and heuristic analysis of known mining behaviors also play a role. For browser-based attacks, blocking JavaScript mining scripts can be effective, although this approach is limited: attackers can obfuscate code, and some websites may not function properly if JavaScript is disabled. With the rise of AI improving attackers' capabilities, cryptojacking remains a persistent threat that requires adaptive defenses~\cite{hong2021survey}.  Digital Forensics and Incident Response (DFIR) can be an effective approach for detection and mitigation. Study~\cite{adeniran2024dissecting} suggests that attackers leverage shared hosting, domain obfuscation, and outdated software to persist over time. These findings underscore the need for continuous monitoring, better transparency in domain registration, and cross-layer detection techniques to counter the evolving infrastructure behind cryptojacking campaigns

\subsubsection{Emerging Challenges}
Cryptojacking has evolved beyond simple browser-based attacks to sophisticated techniques that exploit cloud infrastructure, fileless malware, and memory-resident scripts. Attackers increasingly use stealthy methods such as PowerShell, WMI, and obfuscated JavaScript to evade detection, while mining-as-a-service platforms and third-party software vulnerabilities complicate defenses. Recent research~\cite{chmiel2024evade} has shown that even advanced detection tools can be bypassed by tactics like function renaming, loop unrolling, and modifying WebSocket behaviors, highlighting how quickly attackers adapt. As a result, cryptojacking often goes unnoticed for long periods.

\subsection{Deepfake-Enabled Fraud}

Deepfakes-enabled frauds are Generative AI created content to impersonate individuals or fabricate evidence involving images, audios, or videos~\cite{chesney2019deepfakes}. In recent years, deepfakes have been increasingly used to cause identity fraud, social engineering attacks, mass public deception, and public image tarnishing~\cite{mirsky2021creation}. Deep fake can be active or passive, active deep fake includes Face and voice swapping in audio and videos to impersonate a person in live video calls~\cite{agarwal2020detecting}, causing business-related authentication to be broken than relying on audio and video of a person. The passive attacks include distribution of impersonated images to cause social reputation harm or public deception, such as to cause policy changes and election manipulation~\cite{pawelec2022deepfakes}, see many such in recent election years, as well as celebrity or public personabilities in porn or other places, say memes~\cite {fido2022celebrity,sharma2025role}. Attackers often combine deepfakes with traditional settings to improve the impact of the attack.

\subsubsection{Impact}

Deepfake-enabled fraud impacts the social reputation of the person, resulting in damages that span from monetary~\cite{chesney2019deepfakes}, image tarnishing, to the ability to be elected or hold a position of importance. Effectively, it creates a major distrust among the public about digital communication, where the public can view legitimate media as false.

\subsubsection{Detection, Mitigation and Prevention}

Media forensics, verification, control, and independent trust infrastructure are major techniques to detect, mitigate, and prevent such frauds. Deepfake detection methods utilize various inconsistencies in artifacts~\cite{tolosana2020deepfakes}. Some modern techniques include frequency domain analysis and deep learning  ~\cite{frank2020leveraging,heidari2024deepfake}. There are domain-specific techniques that also exist, for example,  voice biometric systems targeted to distinguish synthetic speech from real audio~\cite{wang2020asvspoof}. Modern trust models include the use of Blockchain and digital watermarking to perform the authenticity of media~\cite{verdoliva2020media}. Generative AI and the proliferation of ready-to-use open source tools are helping this attack evolve, making it hard to detect

\subsubsection{Emerging Challenges}
The increasing realism and accessibility of deepfake tools make fraudulent impersonation more convincing and widely available. Attackers are also blending deepfakes with phishing, vishing, and social engineering to maximize impact, creating complex multi-vector threats~\cite{heidari2024deepfake}. At the same time, gaps in legal and policy frameworks hinder effective regulation, and the erosion of trust in digital media enables the “liar’s dividend,” where authentic evidence can be dismissed as fake~\cite{khan2025unmasking,amerini2025deepfake}. These factors collectively position deepfake-enabled fraud as one of the most dynamic and difficult cybercrime categories to address.

\begin{table}[ht]
\centering
\caption{Cybercrime: Summary of Emerging Challenges}
\label{tab:emerging_challenges}
\begin{tabular}{|p{3cm}|p{9cm}|}
\hline
\textbf{Cybercrime Type} & \textbf{Emerging Challenges} \\
\hline
\textbf{Ransomware} & Increasing sophistication of encryption techniques; use of double/triple extortion (data theft + DDoS + leaks); targeting of critical infrastructure; Ransomware-as-a-Service (RaaS) ecosystems lowering entry barriers. \\
\hline
\textbf{Phishing} & Highly personalized spear-phishing using AI; multi-vector attacks combining email, SMS, and social media; use of deepfake audio/video in vishing; evasion of spam filters via adversarial content generation. \\
\hline
\textbf{Intrusions / APTs} & Advanced persistent threats with stealthy lateral movement; supply-chain infiltration; “living off the land” techniques to avoid detection; exploitation of zero-day vulnerabilities. \\
\hline
\textbf{CSAM} & Increasing use of encrypted platforms; AI-generated synthetic CSAM; distribution via decentralized networks; legal and jurisdictional barriers to enforcement. \\
\hline
\textbf{Cyberbullying} & Use of ephemeral messaging apps to evade detection; AI-generated harassment content (text-to-image/video abuse); psychological impact magnified by algorithm-driven amplification; cross-platform targeting. \\
\hline
\textbf{Cryptojacking} & Targeting of cloud/IoT environments for large-scale mining; polymorphic malware that adapts to avoid detection; economic incentives driving constant innovation; integration with other malware campaigns. \\
\hline
\textbf{Deepfake-Enabled Fraud} & Realism; accessibility of tools; arms race with detection methods; combination with phishing/vishing/social engineering; legal/regulatory gaps; erosion of trust via “liar’s dividend.” \\
\hline
\textbf{Supply Chain Attacks} & Growing exploitation of third-party vendors and software; difficulty in verifying software integrity; attacks on CI/CD pipelines; widespread downstream impact due to single compromise. \\
\hline
\end{tabular}
\end{table}

\subsection{Summary }
We developed a taxonomy of cybercrime encompassing traditional threats such as ransomware, phishing, network intrusions, and child sexual abuse material, as well as emerging challenges like cryptojacking, deepfake-enabled fraud, and supply chain attacks. Each category has distinct characteristics, impacts, and detection strategies. Table~\ref{tab:emerging_challenges} consolidates emerging challenges highlighting how each threat type is evolving, and the systemic difficulties that complicate detection, mitigation, and prevention. In summary, defending against modern cybercrime requires a combination of technical safeguards, policy frameworks, and multidisciplinary collaboration. As threats continue to evolve, continuous research and adaptive security strategies remain critical.

\section{Conclusion}
\label{future}
This paper presents a structured Taxonomy of major cybercrime called STRIKE. The Taxonomy chart classifies the cybercrimes based on the type of attack vectors, operational impact, informational impact, defense and targets. This classification scheme will aid a defender in protecting their network by providing vital attack information.

There is an ongoing new attack manifestation, therefore this taxonomy could be extended to include new categories within each classification. It will provide a defender with the appropriate information to make an educated decision in defending against cyber attacks. Creative approaches to defending attacks will become available, and providing an extensible taxonomy able to capture new defenses is imperative to defense. We believe this taxonomy provides a foundation for the cybersecurity community and provides the ability to continuously grow as attacks and defenses become more sophisticated. In future work, to build a better Defense System, more research can be done to see the applicability of this taxonomy in determining the action space of the attackers.
%
%
%
\bibliographystyle{splncs04}
\bibliography{reference}
%

\end{document}